# Communicating Concurrent Functions

*Bob Diertens*

section Theory of Computer Science, Faculty of Science, University of Amsterdam

*ABSTRACT*

In this article we extend the framework of execution of concurrent functions on different abstract levels from previous work with communication between the concurrent functions. We classify the communications and identify problems that can occur with these communications. We present solutions for the problems based on encapsulation and abstraction to obtain correct behaviours. The result is that communication on a low level of abstraction in the form of shared memory and message passing is dealt with on an higher level of abstraction.

*Keywords:* communicating functions, computational model, concurrency

## 1. Introduction

In [1] we developed a framework of computational models at different levels of abstraction for the concurrent execution of functions. The development of the framework started with the traditional sequential execution model for functions from which a sequential computational model was obtained by abstracting from the details of function call implementation. Further abstraction from the way a function is scheduled for execution led to an abstract computational model that allows for the concurrent execution of functions (Figure 1).

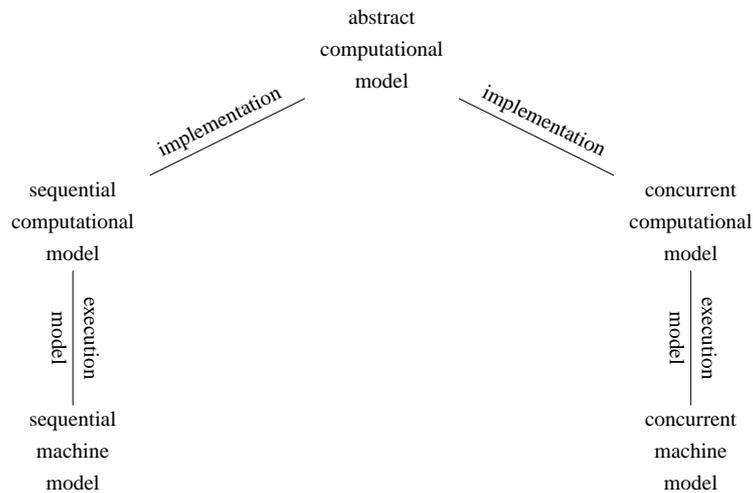

**Figure 1.** Framework of computational models

It showed that with abstraction and relaxing constraints a model for execution of functions can be obtained in which function scheduling plays a key role. This model has as a possible implementation inline scheduling, the original stack-based function execution model the development of the framework started with. But more important, this model allows for the concurrent executions of functions, and therefore it can be used as model for the implementation of concurrent software. The framework of computational models at different levels of abstraction can be used for further development of concurrent computational models that deal with the problems inherent with concurrency.

Concurrent execution of code allows for communication between the parts of code that are executed concurrently. The communication can be done either through models using shared memory or by models using message passing. With message passing models communication is done by exchanging messages and shared memory communication is done by reading from and writing to memory that can be accessed by at least all parties in the communication. In this paper we extend the abstract model for concurrent execution of functions with communication between concurrent functions.

We classify communications between concurrent functions in direct and indirect, and in unidirectional and bidirectional. In the following we describe the different forms of communications and identify the problems that can occur with the communications. We describe solutions for dealing with these problems as abstract as possible as we are dealing with communications on a higher level of abstraction than it is usually the case. The reason for dealing with communications on a high level of abstraction is that we are only concerned with the essentials of the communication and not with the implementation details. Faithful implementations should behave the same as their abstract counterparts when abstracting from the inner details. These implementations should also prevent other concurrent functions from interfering with the inner details. Further, we describe some special cases of communication and how they relate to the forms of communications and the solutions already described.

## 2. Forms of Communication

Communications between concurrent functions can either be direct or indirect. With direct communication one function sends a message and another function receives the message. On the level of abstraction we are working, the combination of the sending of the message and the receiving of the message between concurrent functions can be seen as a single indivisible action.

With indirect communication one function writes a message to a dedicated location and another function reads the message from that location. Indirect communication conversely can not be seen as an indivisible action as parts of the indirect communication may interfere with eachother. We list some problems that can occur with indirect communication.

- If the write and read actions are divisible then these actions can interfere with each other.
- It is possible that a new message is written before the old message is read.
- If the location the messages are written to is accessible by other concurrent functions, interference from these functions is possible as well.

As opposed to unidirectional communications, bidirectional communication can occur between concurrent functions. Bidirectional communication between two functions consists of two unidirectional communications. Bidirectional direct communication can lead to a deadlock when both function want to either write or read a message. Such a problem can be solved by using indirect communication with at least one of the unidirectional communications.

With bidirectional indirect communication interference between the two unidirectional communications is possible when the same location for storing the messages is used. Messages can be overwritten before they are read. If that is not allowed, separate locations for storing the messages should be used for the two unidirectional communications. Of course when it is known that this is never the case sharing the location is an option, but one that is prone to errors by later alterations. If it is allowed, it means that messages do not have to be read. But then write actions from both sides can interfere with each other.

## 3. Solutions to Problems in Communications

The main problem with indirect communication is the interference between the read and write actions of the communication partners. Therefore, we first present a basic locking mechanism that guarantees exclusive access to the location used for storing the message. We then give solutions to the problems with unidirectional and bidirectional indirect communication based on this mechanism.

*3.1 Basic Locking Mechanism*

To avoid problems with using indirect communication between concurrent functions a mechanism is needed that prevents simultanous access to the location used for storing the messages. A possibility for this is the use of a locking mechanism that operates in concurrency with the communicating functions and that has two actions associated with it, a lock action and an unlock action. A lock action gives exclusive access to a particular location, and an unlock action releases the exclusive access (Figure 2).

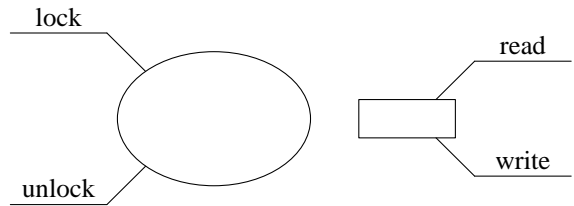

**Figure 2.**  Indirect communication with locking mechanism

A read or a write action may only occur after a lock action and must be followed by an unlock action. This locking mechanism is commonly used in software, where it is implemented with two semaphores using Dekker's algorithm described in [2].

Although this mechanism may work fine in a well disciplined environment, there still are problems with this mechanism since it does not prevent access to the location for storing the messages without using the mechanism. A solution to this is to use a locking mechanism that encapsulates the location. The encapsulation can be established by using a location that is only locally accessible instead of a location that is globally accessible (Figure 3).

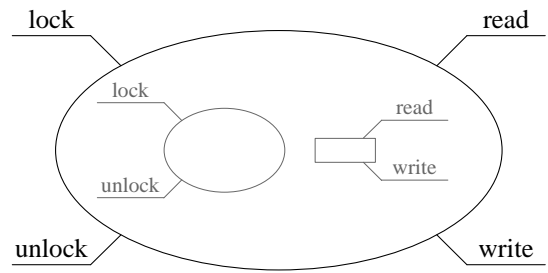

**Figure 3.**  Indirect communication with encapsulated location and locking mechanism

With this mechanism, a read or a write action can only take place when first a lock action has been done by the function that wants to read or write. So for writing a message into the location the sequence of actions lock - write - unlock and for reading a message from the location the sequence lock - read - unlock have to be performed.

*3.2 Unidirectional Indirect Communication*

In this section we describe the implementation of a mechanism for unidirectional indirect communication based on the basic locking mechanism from the previous section. We start with a mechanism that implements the basic communication actions, which we extend to a mechanism that fully implements unidirectional indirect communication.

**3.2.1 Basic Communication Mechanism**
When using the basic locking mechanism writing a message consists of the sequence of actions lock - write - unlock, and reading a message of the sequence lock - read - unlock. From the point of view of the writer and reader of the message these sequences can each be incorporated in a single action. Instead of implementing such actions on the side of the writer and reader, we can implement a mechanisme that provides these actions (Figure 4). That mechanism can at the same time encapsulate the basic locking mechanism to prevent access to it directly.

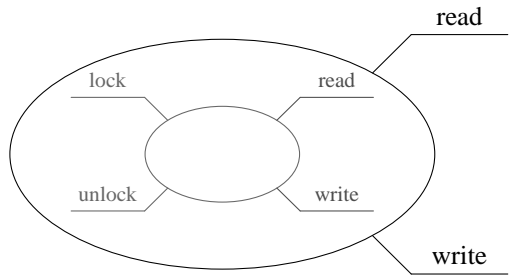

**Figure 4.** Indirect communication with encapsulated locking mechanism

When we abstract from the inner working of this mechanism, it is just another concurrent function with associated read and write actions. The form of communication with this function is unidirectional direct communication. Thus we can conclude that unidirectional indirect communication is built up from two unidirectional direct communications, one between the writer and the mechanism and one between the mechanism and the reader.

**3.2.2 Status Based Communication**

With unidirectional indirect communications only a read may be done when a message first has been written. And writing of the next message may only be done when the previous message has been read. The basic communication mechanism from the previous section does not have these constraints. To enforce these constraints, we have to keep track of the status of the basic communication mechanism and only a write or a read based on the current status.

We add a status flag with associated actions set and get in order to see if a message can be written or that a message has to be read. We encapsulate the status flag and the basic communication mechanism to prevent interference by wrapping them in another mechanism. We can then abstract from the innerworking of this mechanism. We are left with a mechanism that can only do a read action when a message has been written and only a write action when there has been no write yet or the message has been read. We provide the mechanism with an action for checking the status from outside the mechanism.

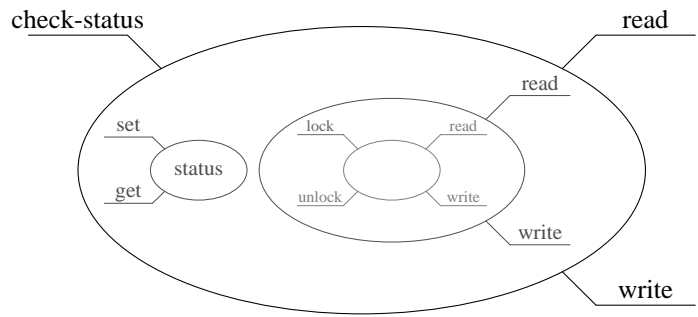

**Figure 5.** Unidirectional indirect communication with encapsulated status flag

*3.3 Bidirectional Indirect Communication*

When bidriectional indirect communication consist of two unidirectional indirect communications there are no additional problems and we solve them as described above. When bidirectional indirect communication is built using only one location for storing the messages in both directions the same problems arise as with unidirectional indirect communication. There is only one additional problem, and that is the possibility of overwriting a message before it can be read due to interference of the communications in the two directions.

The solution to this problem is similar to the solution described above, but it is more complex since now the two directions of communication have to be taken into account. We have to alter the possible number of states to make sure that a message can only be written when there is no message waiting to be read and that a message can only be read by the one side it is intended for.

## 4. Special Cases

In this section we discuss soem special cases of communications and how the mechanism from the previous sections can be used in these cases.

### *4.1 Last Message*

It can be the case that the receiver of the messages is not interested in all messages, but only the most recent one. With unidirectional indirect communication we can relax on the constraint that a message has to be read before another write of a message can take place. The receiver can then ignore messages and the sender does not have to wait for a message to be read before writing the next one. Depending on whether the involved parties want the check on the status of the communication mechanism, an implementation can either use the status based communication mechanim from section 3.2.2 or fall back to the basic communication mechanism from section 3.2.1.

A similar thing can be done for bidirectional indirect communication with sharing of the location for the messages in both directions. But then we have to make sure that it is always possible to read the last message by holding on to the constraint that a message can only be written by a side if there is no message waiting to be read by this side. So a status is still needed and thuis we have to use the status based communication mechanism.

### *4.2 Undirected Communication*

If we dispose of all constraints with bidirectional indirect communication with sharing of the storage location for messages in both directions we get a form of communication that can be considered as undirected. All parties involved can write and read from the location at any time. In this case the communication is not with eachother anymore, but with the storage location. This is what at a low level of abstraction typically is meant by shared memory. It can be implemented with the basic communication mechanism.

In using undirected communication it can be required to generate a new message based on the latest available message. To make sure the new message is based on the latest message and that there has not been written a new message in the meantime, it must be possible to lock the location during the process. A sequence of actions has to be lock - read - generate - write - unlock. So in this case, an implementation has to fall back to the basic locking mechanism from section 3.1.

## 5. Conclusions

We have extended the abstract computational model from our framework of computational models for the concurrent execution of functions with communication between the concurrent functions. For this, we classified the communications between concurrent functions in direct and indirect and in unidirectional and bidirectional. For the different kinds of communication we identified problems that can occur with these communications and we presented solutions for these problems. The solutions are based on encapsulating the area in which the problem occurs together with a mechanism that solves the problem. The encapsulation makes it possible to abstract from the innerworkings. The resulting mechanism can then easily be replaced with something else with the same behaviour.

We defined the basic locking mechanism that we use in our solutions. This mechanism itself is based on the locking mechanism commonly used in software, but through encapsulation and abstraction this is completely concealed.

For unidirectional indirect communication we defined the basic communication mechanism that encapsulates the basic locking mechanism. This basic communication mechanism is constrained into the status based communication that implements unidirectional indirect communication. The constraints are based on the states of the mechanism that indicate whether a read or a write is possible.

Bidirectional indirect communication consisting of two unidirectional indirect communications that share

- 6 -the location for storing the messages in both directions can be implemented using a similar status based communication. But then the solutions is a bit more complex because there are more possible states. Not only is it necessary to indicate whether a read or a write is possible, but also by which party involved in the communication.

Furthermore, we discussed some special cases of communication. These cases can be implemented using one of the defined mechanisms, or similar implementations with the same behaviour, depending on the constraints that have to be enforced on the communications. A particular intriguing case is undirected communication whatat a low level of abstraction is meant by shared memory communication. This can be implemented with the basic communication mechanism or the basic locking locking mechanism when control over the locking mechanism itself is required.

When we abstract from the inner working of the mechanisms, these are just other concurrent functions among the other functions. The form of communication with these mechanisms is unidirectional direct communication. So we can conclude that, through our mechanisms, unidirectional indirect communication is built up from unidirectional direct communications.

# References

[1]  B. Diertens, *Concurrent Models for Function Execution,* section Theory of Computer Science - University of Amsterdam, 2011.

[2]  E.W. Dijkstra, *Cooperating Sequential Processes,* Technological University Eindhoven, 1965.